\documentclass[preprintnumbers,secnumarabic,amsmath,amssymb,nofootinbib,floatfix]{revtex4}
\usepackage{varioref,exscale,latexsym,amsmath,amssymb}
\usepackage{graphicx}
\usepackage{graphicx}% Include figure files
\usepackage{slashed}
\usepackage{dcolumn}% Align table columns on decimal point
\usepackage{bm}% bold math
\usepackage{hyperref}
\usepackage{color}
\usepackage{xcolor}
\newcommand{\beq}{\begin{equation}}
\newcommand{\eeq}{\end{equation}}
\newcommand{\bea}{\begin{eqnarray}}
\newcommand{\eea}{\end{eqnarray}}

\def\lsi{\raise0.3ex\hbox{$<$\kern-0.75em\raise-1.1ex\hbox{$\sim$}}}
\def\gsi{\raise0.3ex\hbox{$>$\kern-0.75em\raise-1.1ex\hbox{$\sim$}}}

\def\beq{\begin{equation}}

\def\eeq{\end{equation}}

\def\beqa{\begin{eqnarray}}

\def\eeqa{\end{eqnarray}}

\begin{document}
\preprint{ACFI-T21-05}

\title{{\bf The Ostrogradsky instability can be overcome by quantum physics }}

\medskip\

\medskip\

\author{John F. Donoghue}
\email{donoghue@physics.umass.edu}
\affiliation{~\\
Department of Physics,
University of Massachusetts\\
Amherst, MA  01003, USA\\
 }

\author{Gabriel Menezes}
\email{gabrielmenezes@ufrrj.br}
\cite{}\affiliation{~ Departamento de F\'{i}sica, Universidade Federal Rural do Rio de Janeiro, 23897-000, Serop\'{e}dica, RJ, Brazil \\
 }

\begin{abstract}
In theories with higher time derivatives, the Hamiltonian analysis of Ostrogradsky predicts an instability. However, this Hamiltonian treatment does not correspond the way that these theories are treated in quantum field theory, and the instability may be avoided in at least some cases. We present a very simple model which illustrates these features.
\end{abstract}
%\vspace{0.2 in}
%\end{titlepage}
%\setcounter{page}{0}
%\newpage
\maketitle
%\documentstyle[12pt,epsfig]{article}
%\documentstyle[12pt,epsf,epsfig]{article}

%%%%%%%%%%%%%%%%%%%%%%%%%
\section{Introduction}
In 1850, Ostrogradsky analysed Lagrangians which contained higher time derivatives and showed that such theories are classically unstable \cite{Ostrogradsky:1850fid, Woodard:2015zca}. Classical instability need not imply quantum instability. A counter-example is the theory of the Dirac equation, where the classical Hamiltonian has unbounded negative energies while the quantized Dirac field is stable with positive energies. For higher derivative theories there remains much debate about stability, positive energies, unitarity and causality. We here present a simple model representative of a class of theories and show that the Ostrogradsy instability is not present, as well as showing that the massive excitation carries positive energy and that the classical limit is normal. We have elsewhere demonstrated unitarity for this class of theories \cite{Donoghue:2019fcb}, and have also discussed causality \cite{Donoghue:2019ecz, Donoghue:2020mdd, toappear}. We also review these elements below.

The work on quantum field theories with higher derivatives goes back to Lee and Wick \cite{Lee:1969fy, Lee:1969fz, Lee:1970iw} in the 1960's and our discussion incorporates elements of this past work. Much of the present interest arises from quadratic gravity, a renormalizeable theory where the Lagrangian contains squares of the curvatures as well as the Einstein term linear in the curvature \cite{Stelle:1976gc, Julve:1978xn, Fradkin:1981hx, Tomboulis, Salvio:2018crh, Donoghue:2018izj, Einhorn, Strumia, DM, Holdom, Mannheim, Shapiro, Narain, Anselmi}. Because the curvatures tensors are second order in the derivatives of the metric, this becomes a theory with four derivatives in the Lagrangian. Quadratic gravity falls in the category which we are discussing here. However, it is a far more complicated theory. We hope that our simple model here illustrates  the key features for this class of theories.

\section{The model}

Consider a ``normal'' theory (i.e. without higher derivatives) of a complex scalar field $\chi$ coupled to a real scalar $\phi$ with the Lagrangian
\beq
{\cal L}= {\cal L}_\chi +{\cal L}_\phi -g\phi \chi^\dagger \chi
\eeq
with
\beqa
{\cal L}_\phi &=& \frac12 \left[\partial_\mu \phi \partial^\mu \phi - m^2 \phi \phi \right]  \nonumber \\
{\cal L}_\chi &=& \partial_\mu \chi^\dagger \partial^\mu \chi - m_\chi^2 \chi^\dagger \chi - \lambda (\chi^\dagger \chi)^2 \ \ .
\eeqa
The relative masses here are not particulary important, but we will present results for $m^2< m_\chi^2$. While there is no symmetry which can force $m  =0$, we can nevertheless envision a situation where the renormalized mass of the $\phi$ field is negligibly small. This then would be a scalar model for a charged field $\chi$ interacting with a scalar photon $\phi$. The exchange of the $\phi $ would yield a Coulomb - like potential. If the mass were zero, the $\phi $ field would satisfy the classical wave equation. So we invite the reader to consider this as a scalar model for QED, or perhaps as a scalar model for the gravitational interaction of a massive $\chi$ field. Note that as a writing style, we will refer to theories with only two time derivatives as ``normal'', so that we do not have to regularly specify that there are only two derivatives.

Into this bucolic setting we now introduce a troublesome term with higher derivatives
\beq
{\cal L}_\phi \to {\cal L}_{hd}
\eeq
with
\beq\label{fullmodel}
{\cal L}_{hd}= \frac12 \left[\partial_\mu \phi \partial^\mu \phi - m^2 \phi \phi - \frac1{M^2} \Box\phi \Box\phi \right] \ \ .
\eeq
Here $M$ is a large mass, very much larger than $m$ and $m_\chi$. We invite the reader to consider $M$ as the Planck mass - far beyond the range of ordinary experiments.

We know from work on effective field theories that if we treat this new term as a perturbation, it would have a negligible effect at low energies, and would not change the classical wave equation if $m=0$. However if we treat this as a fundamental theory, the analysis of Ostrogradsky would say that the theory has an instability which renders the theory non-viable even at low energy. This is the simple model which we wish to analyze.

\section{The low energy/classical limit}

Let us define the quantum theory by using path integrals. We first focus on the path integral over the $\phi$ field
\beq
Z_\phi [\chi] = \int [d\phi ] e^{i \int d^4x [{\cal L}_{hd} - g\phi \chi^\dagger \chi]} .
\eeq
We can now manipulate this a bit. We introduce an auxiliary field $\eta$ which, when you integrate it out, reproduces the same Lagrangian. This is
\beq
{\cal L}(\phi, \eta) =  \frac12 \partial_\mu \phi \partial^\mu \phi  - \eta \Box \phi + \frac12 M^2 \eta^2 -g\phi \chi^\dagger \chi \ \ .
\eeq
This results in
\beq
Z_\phi [\chi] = \int [d\phi ] [d\eta] e^{i \int d^4x [{\cal L}(\phi,\eta)]} \ \ .
\eeq
As a next step we can define a new field by $\phi(x) = a(x) -\eta(x)$ replacing the field $\phi $ by this combination. The Lagrangian then completely separates and becomes
\beqa\label{separated}
{\cal L}(a, \eta) &=&  \left[\frac12 \partial_\mu a \partial^\mu a -g a \chi^\dagger \chi\right]  \nonumber \\
&-&~  \left[\frac12 \partial_\mu \eta \partial^\mu \eta -\frac12 M^2 \eta^2 -g \eta \chi^\dagger \chi\right]
\eeqa
In summary, we have transformed the original theory exactly to
\beqa
Z_\phi [\chi] &=& \int [da] e^{i \int d^4 x \left[\frac12 \partial_\mu a \partial^\mu a -g a \chi^\dagger \chi\right] } \nonumber \\
  &\times& \int [d\eta] e^{-i \int d^4 x \left[\frac12 \partial_\mu \eta \partial^\mu \eta -\frac12 M^2 \eta^2 -g \eta \chi^\dagger \chi \right]}  \nonumber \\
  &=& Z_a \times Z_\eta
\eeqa
The first path integral, over $a(x)$, is just the original normal theory, with $\phi(x)$ replaced by $a(x)$. The second path integral, over $\eta (x)$, is the complex conjugate of a normal massive theory.  The remnant of the original higher derivative term is the $-i$ instead of $+i$ in the second path integral.

We defer any interpretation to later. For now, let us just calculate the path integral over $\eta$. This is a Gaussian integral and is perfectly well defined. The result is simply the complex conjugate of the usual Gaussian integral. To see this explicitly, we add an real infinitesimal factor of $-\epsilon \int d^4x  \phi^2$ to the exponent to make the result well behaved for large fields. Then we complete the square using
\beq
\eta'(x) = \eta(x) - \int d^4 x iD_{-F}(x-y) ~\chi^\dagger(y) \chi(y)
\eeq
with
\beq
iD_{-F}(x-y) = \int \frac{d^4k}{(2\pi)^4} \frac{-i}{k^2- M^2-i\epsilon} \ \ .
\eeq
This propagator is the complex conjugate of the usual Feynman propagator, changing the sign in the numerator and also the sign of the $i\epsilon$ term in the denominator. The integral over $\eta'$ yields
\beq
Z_\eta = N e^{\int d^4x d^4y \frac12 g\chi^\dagger(x)\chi(x) ~i D_{-F}(x-y) ~g\chi^\dagger (y)\chi(y)} \ \ .
\eeq
At low energy, the interaction becomes local and we obtain
\beq
Z_\eta = N e^{i\int d^4x  \frac{ g^2}{2M^2}[\chi^\dagger(x)\chi(x)]^2 }  \ \ .
\eeq
This is just a shift in the quartic interaction of the $\chi$ field, with
\beq
\lambda \to \lambda' = \lambda - \frac{ g^2}{2M^2}  \ \ .
\eeq
The minus sign in the new contribution is the remnant of the use of $\exp (-iS)$ in the path integral. However, for a large mass $M$, this will not change the sign of $\lambda'$.

The low energy limit of this theory, quantized using path integrals, is then perfectly normal. The resulting classical theory for small or vanishing $m$ is then also unchanged. We colloquially refer to the classical limit as taking $\hbar \to 0$. However in fact $\hbar$ is a fixed constant, and the classical regime is that with kinematics such that $\hbar$ effects are unimportant. We will see that at high energy and short wavelengths $\hbar$ effects are crucial. In this theory the classical limit involves wavelengths much larger than the Compton wavelength of the $\chi$ field, much like in usual QED.

The most appropriate interpretation of the $\eta$ path integral is as the time-reversed version of a regular path integral. Time-reversal is an anti-unitary operation, involving complex conjugation. The Lagrangian itself is time-reversal invariant but in the path integral $\exp (iS)$ changes to $\exp (-iS)$. Within the path integral, this change is manifest most importantly in the $i\epsilon$ in the propagators. These define the arrow of causality \cite{Donoghue:2019ecz, Donoghue:2020mdd}- telling us what is the past lightcone and what is the future. We will see that when we decompose the propagator into time ordered factors, the usual $i\epsilon$ tells us that positive energies propagate forward in time. Changing the sign on $i\epsilon$ leads to propagation of positive energies backwards in time. This is described explicitly in the following section.

\section{High energy}

In this section, we show how the coupling to the $\chi$ fields makes the heavy particle decay, that positive energy is needed to excite this resonance and we further demonstrate the ``backwards in time'' behavior of the resonance.

While one often starts the analysis of a theory in the free-field limit with no interactions, here it is important to include the effect of interactions in order to properly understand the spectrum of the theory. In this regard, it is more similar to the analysis of the electroweak theory, where the interaction with the Higgs boson is included from the start in order to get the spectrum correct. In our case here, the coupling to the $\chi$ fields is required to provide information on the decay width which is crucial for understanding the spectrum.

Consider the $\phi$ propagator in the original basis, before any field redefinitions have been performed. Including the vacuum polarization, this has the form
\beq
iD(q^2) =\frac{i}{q^2 -m^2 +i\epsilon - \frac{q^4}{M^2} +\Sigma({q}) }  \ \ .
\eeq
The one loop vacuum polarization has a divergent piece which goes into the renormalization of $m^2$. As noted above for convenience we will choose the renormalized value of $m^2$ to vanish, in which case the finite part of the vacuum polarization is
\beq
\Sigma_f (q) = -\frac{g^2}{32\pi^2}\int_{0}^{1} dx \log\left[\frac{m^2_\chi - x(1-x)(q^2+i\epsilon)}{m_\chi^2}\right] \ \ .
\eeq
Beyond $q^2 = 4m_\chi^2$ there will be an imaginary part of the vacuum polarization, which for our purposes is the most important feature. At high $q^2$, where we apply this, we have the result
\beq
\Sigma(q) \sim \frac{g^2}{32\pi^2} \left[-\log\left(\frac{-q^2-i\epsilon}{m_\chi^2}\right)+2\right]
\eeq
such that
\beq
{\rm Im}\Sigma \sim \frac{g^2}{32\pi}  \equiv \gamma \ \ .
\eeq

With this result we can look for the high mass pole. It is found at
\beq
q^2 = \widetilde{M}^2 = \bar{M}^2 +i \gamma
\eeq
where the real part of the mass is found to be
\beq
\bar{M}^2 = {\rm Re}\widetilde{M}^2  \sim M^2 - {\rm Re} \Sigma(M^2)
\eeq
to first order in $g^2$. In the neighborhood of this pole we use $q^2 = \bar{M}^2 +(q^2-\bar{M}^2)$ to find the approximate form
\beqa
iD(q) &\sim& \frac{i}{q^2 - \frac{q^4}{M^2} + {\rm Re}\Sigma +i \gamma }  \nonumber \\
      &\sim& \frac{-i}{q^2 -\bar{M}^2 -i \gamma}  \ \ .
\eeqa
The important thing to notice here is that there are two minus sign differences from a normal resonance. The $-i$ in the numerator and the $-i\gamma$ in the denominator are both of opposite signs from usual resonances. These combined sign differences will lead to the eventual identification of this as the time-reversed version of a usual propagator.

\begin{figure}[htb]
\begin{center}
\includegraphics[height=80mm,width=80mm]{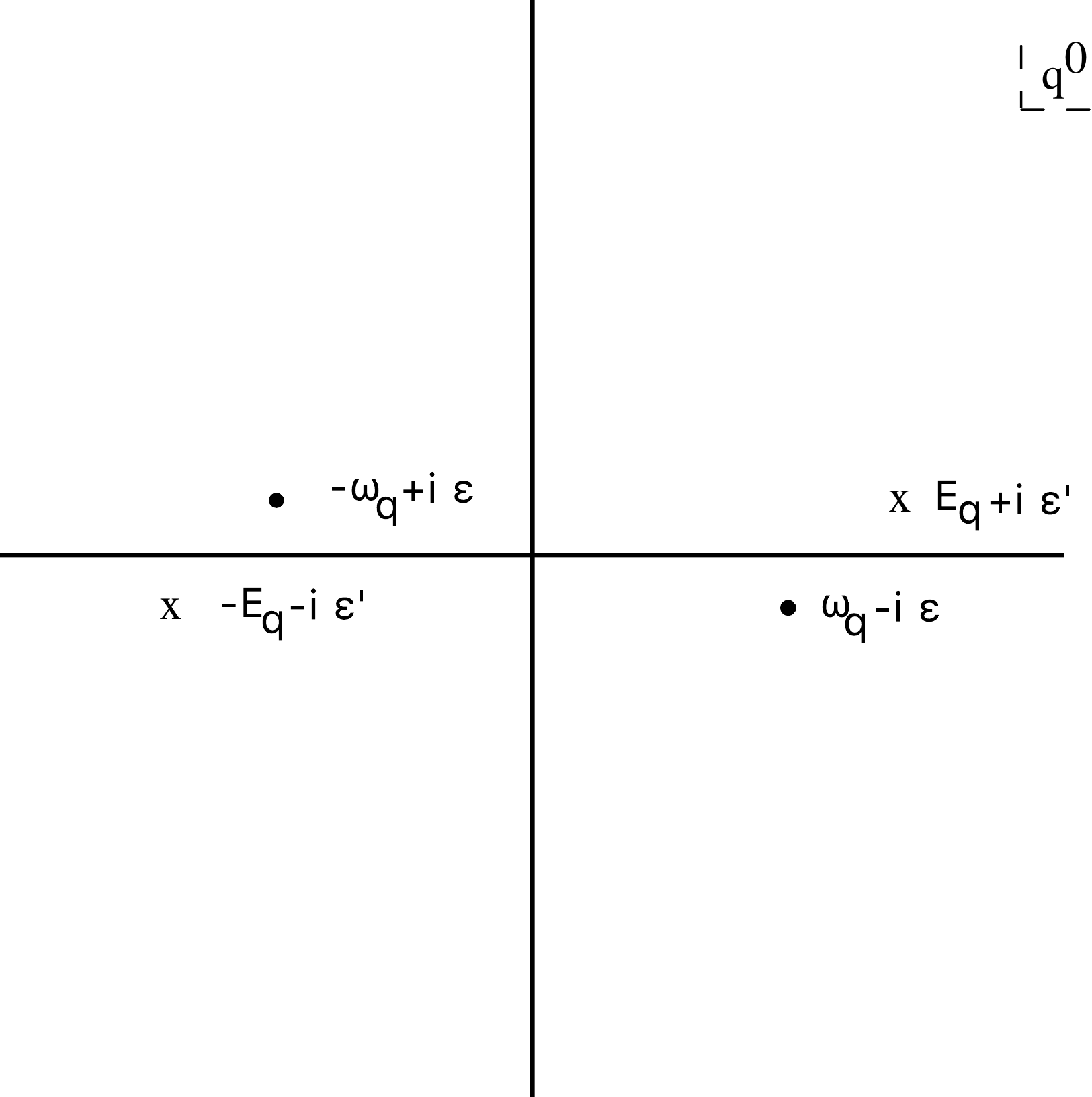}
\caption{The location of poles in the complex $q_0$ plane for the Feynman propagator. }
\label{contour}
\end{center}
\end{figure}

We can see that this propagator corresponds to exponential decay rather than exponential growth by writing it in time ordered form,
\beq
D(t, \vec{x}) = \Theta (t)D_{\textrm{for}} (x) + \Theta (-t) D_{\textrm{back}}(x)
\eeq
with $x_0=t$.
The poles in the complex $q_0$ plane are shown in Fig. \ref{contour}. There is the massless pole at
\beq
q_0^2 - \vec{q}^2 + i\epsilon =0
\eeq
which corresponds to $q_0 = \pm(\omega_q - i\epsilon)$ with $\omega_q=|\vec{q}|$.
There are massive poles at
\beq
q_0^2 - \vec{q}^2 - m_r^2 -i\gamma = 0
\eeq
or
\beq
q_0 = \pm \sqrt{E_q^2 +i\gamma} \sim \pm\left[E_q +i
\frac{\gamma}{2E_q}\right]
\eeq
with $E_q = \sqrt{\vec{q}^2+m_r^2}$. When $t>0$, we close the contour in the lower half plane.
This yields the forward propagator
\beq
D_{\textrm{for}}(t,\vec{x}) = -i\int \frac{d^3q}{(2\pi)^3 }\left[\frac{e^{-i(\omega_q t -\vec{q}\cdot\vec{x})}}{2\omega_q } - \frac{e^{i(E_q t -\vec{q}\cdot\vec{x})}}{2(E_q +i\frac{\gamma}{2E_q})} e^{-\frac{\gamma t}{2E_q}}\right]
\eeq
which shows the decaying exponential for the massive term, with the identification
\beq\label{width}
\gamma = m_r\Gamma \ \ .
\eeq
The term describing propagation backwards in time is obtained for $t<0$ by closing in the upper half plane, with the result
\beq
D_{\textrm{back}}(t,\vec{x}) = -i\int \frac{d^3q}{(2\pi)^3 }\left[\frac{e^{i(\omega_q t -\vec{q}\cdot\vec{x})}}{2\omega_q } - \frac{e^{-i(E_q t -\vec{q}\cdot\vec{x})}}{2(E_q +i\frac{\gamma}{2E_q})} e^{-\frac{\gamma |t|}{2E_q}}\right]
\eeq
Again we see exponential decay. The other notable feature is that the direction of energy flow is reversed for the high mass resonance. Whereas the normal massless pole propagates positive energy forward in time, the high mass resonance propagates it backwards in time.

We have elsewhere proposed calling the high mass resonance in this type of theory a {\em Merlin mode} \cite{Donoghue:2019ecz}, named after the wizard in the Arthurian tales who ages backwards in time. This distinguishes it from the more generic phrasing of ``ghost'', which is applied to any field with the minus sign in the numerator of the propagator. For example, Faddeev-Popov ghosts have a negative sign in the numerator but carry the usual $i\epsilon$ in the denominator. Here there is the extra change of sign in the denominator $ -i\gamma$, which is crucial in making this propagator the time-reversed version of a usual resonance propagator.

\begin{figure}[htb]
\begin{center}
\includegraphics[height=80mm,width=80mm]{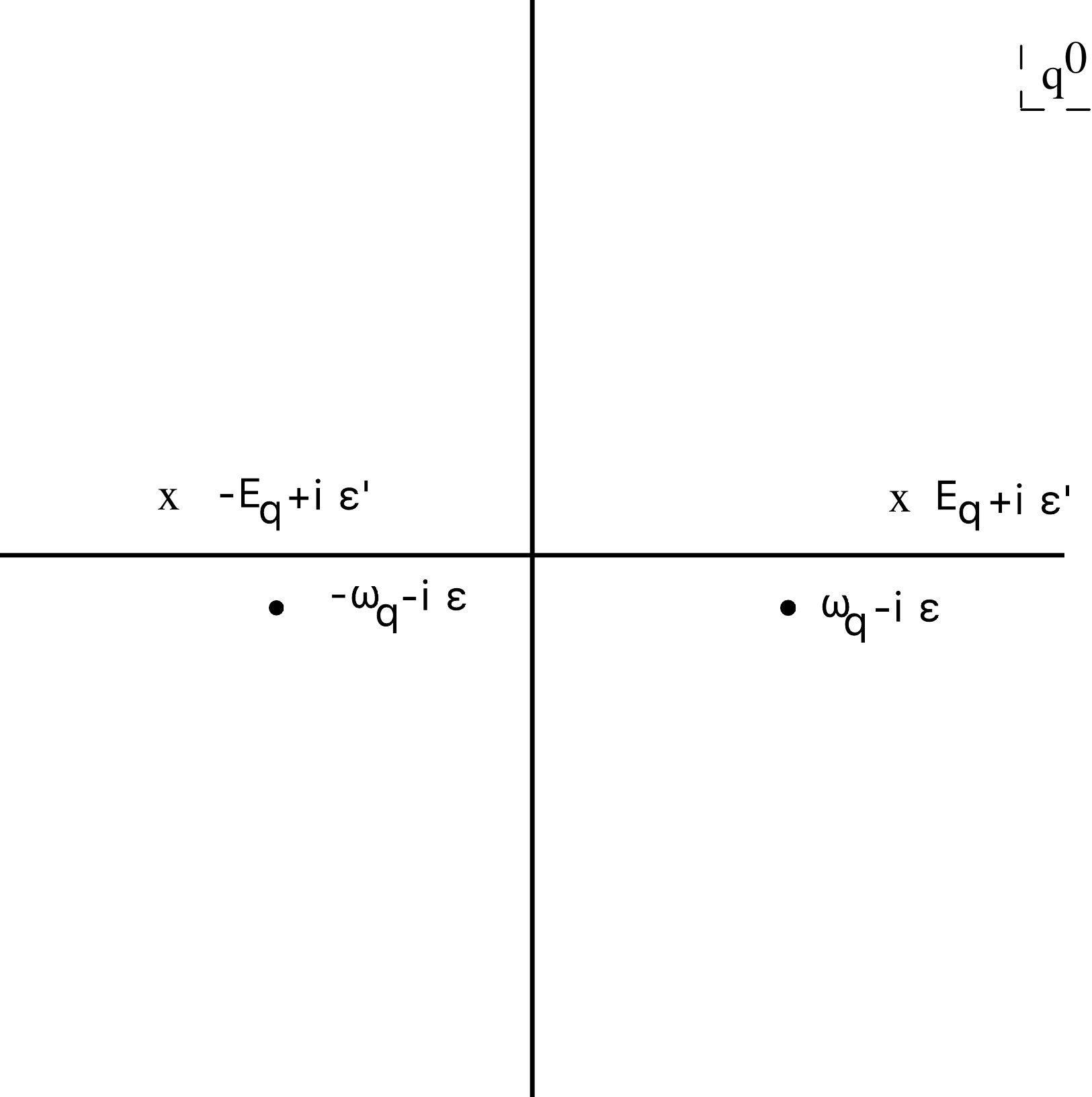}
\caption{The location of poles in the complex $q_0$ plane for the retarded Green function.}
\label{contour2}
\end{center}
\end{figure}

This interpretation is reinforced by calculating the Green function with retarded boundary conditions. The loop integrals for the $\chi$ fields going into the vacuum polarization need to be calculated using the in-in formalism, as in Ref.~\cite{Donoghue:2014yha}. The result is the same functional dependence, but with a different $i\epsilon $ prescription, such that the logarithm is
\beq
\log\left(-\left[(q_0+i\epsilon)^2-\vec{q}^2\right]\right)= \log\left(-q^2 -i\epsilon q_0\right)
= \log|q^2| -i\pi \theta(q^2)\left(\theta(q_0)-\theta(-q_0)\right)   \ \ .
\eeq
This shifts the location of the poles to the positions indicated in Fig. \ref{contour2}. For $t>0$ we pick up the usual
massless poles
\beq
D_{\textrm{ret}}(t>0,\vec{x}) = D_{\textrm{ret}}^{(0)}(t>0,\vec{x})
\eeq
However, even with these boundary conditions, the Merlin resonance gives a contribution for $t<0$,
\beq
D_{\textrm{ret}}(t<0,\vec{x})\equiv D_{\textrm{ret}}^<(t,\vec{x})
=  i\int \frac{d^3q}{(2\pi)^3 }\left[ \frac{e^{-i(E_q t -\vec{q}\cdot\vec{x})}}{2(E_q +i\frac{\gamma}{2E_q})} e^{-\frac{\gamma |t|}{2E_q}}- \frac{e^{i(E_q t -\vec{q}\cdot\vec{x})}}{2(E_q -i\frac{\gamma}{2E_q})} e^{-\frac{\gamma |t|}{2E_q}}\right]  \ \ .
\eeq
This also contains decaying exponentials. If we choose to use this as a Green function giving the response to an external source,
it would correspond to the propagation of the effect backwards in time. This is related to the microcausality violation on scales of order of the resonance width.

The width corresponds to the decay into two on-shell $\chi$ particles. These carry positive energy, so that the resonance also corresponds to positive energy. We can also see that this resonance requires positive energy in order to be produced by the same reasoning. It is seen as an s-channel resonance in $\chi \bar{\chi} \to \chi \bar{\chi}$. The amplitude for the process is
\beq
{\cal M} = g^2 D(q) \sim \frac{-g^2}{q^2 -\bar{M}^2 -i \gamma} \ \ .
\eeq
When squared $|{\cal M}|^2$ has the same form as a usual resonance, so this yields the characteristic Breit-Wigner shape. The incoming $\chi $ fields carry positive energy and one needs a large positive energy to produce the resonance.

With higher derivative theories, there is no guarantee that all quantization methods will yield the same result. The equivalence of various approaches to quantization has been demonstrated only for normal theories. We have used path integral quantization because it is exceptionally clear in this case. However, there are four canonical quantization schemes which we know of which also yield positive energies for the ghost field \cite{Lee:1969fy, Salvio:2015gsi, Raidal:2016wop, Bender:2007wu}. Each requires some modification to traditional canonical quantization. The earliest was due to Lee and Wick in the 1960's where they proposed a higher derivative theory for a finite version of QED \cite{Lee:1969fy, Lee:1969fz, Lee:1970iw} . Their approach was to treat the Pauli-Villars regulator as a dynamical field. The minus sign between the normal propagator and the Pauli-Villars field then becomes the essential complication. They used what they called an ``indefinite metric'' scheme, which modifies the canonical commutation relations. The result is a massive field with positive energy. While our path integral analysis does not rely on the specifics of any canonical quantization scheme, the fact that such schemes exist is welcome.

The high energy structure of this theory is intrinsically quantum. The decay width is crucial for understanding the nature of this resonance. Again we note that while we often refer to the classical world as taking $\hbar \to 0$, in nature $\hbar$ is a constant and the width is a quantum effect. While the exact magnitude of the width is not important, one cannot neglect its effect. Taking the $\hbar \to 0$ version of the propagator functions is not physically sensible.

\section{What would Ostrogradsky say?}

The basic point to be noticed is that the Ostrogradsky construction has no resemblance to quantization via path integrals.

Ostrogradsky's analysis of the higher derivative Lagrangian of Eq. \ref{fullmodel} starts by noting that with extra two time derivatives there are extra degrees of freedom associated with the Lagrangian, and this requires two canonical coordinates and the two associated canonical momentum. His choices for the coordinates are
\beqa
\phi_1 &=& \phi  \nonumber \\
\phi_2 &=& \dot{\phi}
\eeqa
and for the momenta
\beqa\label{momenta}
\pi_1 &=& \frac{\partial {\cal L}}{\partial \dot{\phi}} -\frac{d}{dt}\frac{\partial {\cal L}}{\partial \ddot{\phi}} = \left(\frac{\Box +M^2}{M^2}\right)\dot{\phi} \nonumber \\
\pi_2 &=& \frac{\partial {\cal L}}{\partial \ddot{\phi}}= - \frac{\Box}{M^2}\phi \ \ .
\eeqa
The Hamiltonian is formed by
\beq
{\cal H}(\phi_1,\phi_2,\pi_1,\pi_2)= \pi_1 \dot{\phi_1} +\pi_2 \dot{\phi_2} -{\cal L}  \ \ .
\eeq
In writing this Hamiltonian, we must eliminate $\ddot{\phi}$ in terms of the canonical coordinates and momenta. This is accomplished by using the second line of Eq. \ref{momenta} to write
\beq
\ddot{\phi} = \mathbf{\nabla}^2 \phi -M^2\pi_2   \ \ .
\eeq
The resulting Hamiltonian is
\beq\label{Hamiltonian}
{\cal H} = \pi_1 \phi_2 + \pi_2 \left(\nabla^2 \phi -M^2\pi_2  \right) - {\cal L}(\phi_1,\phi_2, \nabla^2 \phi -M^2\pi_2 ) \ \ .
\eeq
The initial choices of coordinates are then compatible with Hamilton equations
\beqa
\dot{\phi_1} &=& \frac{\partial {\cal H}}{\partial \pi_1}  \nonumber \\
\dot{\phi_2} &=& \frac{\partial {\cal H}}{\partial \pi_2}
\eeqa
and with some effort the Hamilton equation
\beq
\dot{\pi_1} = -\frac{\partial {\cal H}}{\partial \phi_1}
\eeq
can be shown to be equivalent to the Euler-Lagrange equations.

The Ostrogradsky instability is seen in the first term of the Hamiltonian of Eq. \ref{Hamiltonian}. The canonical momentum $\pi_1$ appears linearly, and there is no other factor of $\pi_1$ in the remainder of the Hamiltonian. This implies that the Hamiltonian is not positive definite, and there is no barrier to making the Hamiltonian negative. One does not need energies of order $M$ in order to trigger the instability in this analysis.

To emphasize that the Ostrogradsky construction is not the one relevant for quantum physics, we present the following heuristic version of
Hamiltonian quantization, related to the indefinite metric quantization schemes of Refs. \cite{Lee:1969fy, Salvio:2015gsi, Raidal:2016wop}. We emphasize in advance that this presentation does not do justice to the care taken by those authors, but it does capture how quantization is different from the Ostrogradsky method. If one starts with the separated form for the Lagrangian given in Eq. \ref{separated}
we would define the $\eta$ canonical momentum by
\beq
\pi_\eta = \frac{\partial {\cal L}}{\partial \dot{\eta}} = - \dot{\eta}
\eeq
which has the opposite sign from usual. Imposing the equal-time quantization conditions
\beq
[\eta(x,t), \pi_\eta(x',t) ] =i\hbar \delta^3(x-x')
\eeq
then actually implies the negative of the usual rule, i.e.
\beq
[\eta(x,t), \dot{\eta}(x',t) ] = -i\hbar \delta^3(x-x')  \ \ .
\eeq
(We note that, much like in the path integral analysis, this is the complex conjugate of the usual relation.) To solve this with the usual field decomposition, one would then apply the negative of the usual commutator for the creation operators, i.e.
\beq
[a(p) , a^\dagger(p')] =- \delta^3(p-p') \ \ .
\eeq
If we do this, then the $\eta$ Hamiltonian which emerges from the Lagrangian of Eq. \ref{separated}, i.e
 \beq
 H_\eta = - \int \frac{d^3p}{(2\pi)^3}  \sqrt{p^2+M^2} ~a^\dagger (p) a(p)
 \eeq
actually has positive energy states defined from $|p'\rangle = a^\dagger |0\rangle $.  While Refs. \cite{Lee:1969fy, Salvio:2015gsi, Raidal:2016wop} provide more analysis to be convincing that such constructions are sensible, they do involve changing the commutation relations. As far as Ostrogradsky is concerned however, the main thing to note is that the choice of coordinates and momenta is different. The distinction is in the use of Hamilton's equations. Ostrogradsky has chosen the coordinates to reproduce Hamilton's equations. Quantum physics does not require these. The quantum choice of coordinates is made to produce positive energy states - i.e. quanta. The two choices lead to different Hamiltonians.

We have seen that the Ostrogradsky assignment of $\phi_i $ and $\pi_i$ is not equivalent to the path integral construction of the quantum theory, or even to the canonical methods such as the Lee-Wick indefinite metric quantization. Quantum physics does not use the Ostrogradsky Hamiltonian as it starting point.

\section{Causality and Unitarity}

Despite our analysis of the disconnect between Ostrogradsky and quantum phyics, we know that something has to go wrong in higher derivative theories. Axiomatic field theorists tell us that propagators cannot fall faster than $1/k^2$. This follows from the Kallen-Lehmann representation\cite{Kallen:1952zz, Lehmann:1954xi}
\beq
D_F(k)= \int \frac{d^4k}{(2\pi)^4} e^{ik\cdot x}\langle 0|T \phi(x)\phi(0)|0\rangle = \frac1{\pi}\int ds \frac{\rho(s)}{k^2-s+i\epsilon}
\eeq
where $\rho(s) $ is a positive definite spectral function. If $\rho$ is never negative, the high energy limit has the form
\beq
D_F(k) \sim \frac1{k^2}  \frac1{\pi}\int ds \rho(s) \ \ .
\eeq
If $\rho$ is positive definite, this can never vanish. In higher derivative theories the propagator falls asymptotically as $1/k^4$. Therefore at least one of the axioms which goes into this theorem must be violated.

For our simple theory, the defect is in microcausality. It has been known since the time of Lee-Wick and Coleman \cite{Coleman} that these theories violate causality. For a clear modern exposition, see the work of Grinstein, O'Connell and Wise \cite{Grinstein:2008bg}. The violation is evident from the factor of $-i\gamma$ in the propagator for the Merlin mode, and from the interpretation of this mode as propagating backwards in time. We have written sufficiently on this topic elsewhere \cite{Donoghue:2019ecz, Donoghue:2020mdd, toappear} about this feature that we do not need to repeat that analysis here.

However, unitarity survives intact. We have presented a formal proof of this elsewhere \cite{Donoghue:2019fcb} (see also Lee and Wick \cite{Lee:1969fy}). However the rationale is quite simple to understand. Veltman \cite{Veltman:1963} has shown that the states which appear in the unitarity sum are only the stable states of the theory. The unstable states of the theory are not included in the asymptotic spectrum. With the heavy unstable Merlin mode, the result is the same. The unitarity sum includes only the decay products, which in this case are the $\chi$ fields. One does not include the Merlin mode in the unitarity sum, and hence one is not bothered by the unusual minus signs which appear in the analysis of this state.

While there is no need to repeat the formal proof here, there is a simple and explicit example of how unitarity is manifest in a way which reflects on our treatment of the spectrum above. The s-channel reaction $\chi\bar{\chi} \to \chi\bar{\chi}  $ excites the Merlin resonance. The S-wave partial wave amplitude is
\beq
T_0 (s) = \frac{g^2}{32\pi} D(s)
\eeq
where $D(s)$ is the propagator. Including the fact that near the resonance the width is given by
\beq
\gamma = \frac{g^2}{32\pi}    \ \ ,
\eeq
this then has the near-resonance form
\beq
T_0=\frac{-g^2}{32\pi}\frac1{q^2 -\bar{M}^2 -i \frac{g^2}{32\pi}}
\eeq
which satisfies elastic unitarity
\beq
{\rm Im} T_0 = |T_0|^2  \ \ .
\eeq
Again it is the correlation between the two sign changes characteristic of the Merlin mode which allows unitarity to be satisfied. The asymptotic states here are the $\chi$ fields, as in the Veltman analysis.

However, the existence of the Merlin mode may require further changes to field theory practices at higher orders. For example, Lee and Wick first showed \cite{Lee:1969fy}, and we confirmed in our analysis, that at the two loop order one needs a modification of the contour integral in order to reproduce the discontinuity which is calculated using the Cutkosky rules. The work of Grinstein, O'Connell and Wise \cite{Grinstein:2008bg} contains an explicit example of how this Lee-Wick contour works. At higher order there may be further modifications, or potential problems \cite{Cutkosky:1969fq}. It remains for future investigations to better understand the field theory of these theories.

\section{Summary}

We have displayed a simple higher derivative model whose path integral quantization avoids the Ostrogradsky instability. The main features are the positive energy of the states, and the decay of the ghost field which removes it from the asymptotic spectrum.

The toy model has features which tell us where to look for problems in higher derivative theories of a type which include both the toy model and quadratic gravity. For this class of theories, the problems are not negative energies, nor the Ostrogradsky instability, nor the classical limit. However, there still must be some problem, because such theories do not satisfy all the properties of standard quantum field theories. In this analysis, the problem is microcausality. The heavy ghost field becomes a Merlin mode with differing signs in the numerator and the denominator of the propagator. These two minus signs indicate that the propagation is the T-reversal of a usual resonance.  This leads to ``dueling arrows of causality'', which appears to be the most unusual feature of such theories. If the Merlin particle is heavy enough, and its lifetime small enough, this appears to be compatible with experiment. Further work is needed to better understand the full quantum field theory of these theories. Perhaps a lattice simulation would be useful in order to provide a non-perturbative study. While higher derivative theories have unusual features, perhaps some of them can still lead to reasonable physical theories. 

\section*{Acknowledgements} We would like to thank the following for useful comments or discussions: Simon Caron Huot, G. Dvali, G. 't Hooft, P. Mannheim, A. Salvio, R. Percacci, I. Shapiro, and K Stelle,  and most particularly Bob Holdom and Richard Woodard. The work of JFD has been partially supported by the US National Science Foundation under grant NSF-PHY18-20675. The work of GM has been partially supported by  Conselho Nacional de Desenvolvimento Cient\'ifico e Tecnol\'ogico - CNPq under grant 310291/2018-6 and Funda\c{c}\~ao Carlos Chagas Filho de Amparo \`a Pesquisa do Estado do Rio de Janeiro - FAPERJ under grant E-26/202.725/2018.

\section*{Appendix - on effective field theory}

When working at low energy, heavy fields can be integrated out and we only need to work with those degrees of freedom which are active at the low energy scale. The resulting effective Lagrangian can be expanded in a derivative expansion, and so that the low energy world {\em always} contains higher derivative interactions describing interactions from the full theory which are suppressed by powers of the heavy masses. In this sense, all of our theories are effective field theories with higher derivative corrections at low enough energy. A simple example is the effective Lagrangian for the photon at energies well below the electron mass, having integrated out the electron via the vacuum polarization diagram. The result is
\beq\label{QED}
{\cal L}_{eff} = -\frac14 F_{\mu\nu}F^{\mu\nu} + \frac{\alpha}{60\pi m_e^2}F_{\mu\nu}\Box F^{\mu\nu} +~... \ \ .
\eeq
The second interaction yields a contribution to the Lamb shift.

It is important to emphasize that higher derivatives in the effective Lagrangian do not cause any instability problems. Indeed the QED case is the ``experimental'' verification of this, as QED is perfectly stable at low energies. The Ostrogradsky analysis is not relevant for effective field theories. One quantizes the effective field theory using only the lowest order Lagrangian, and treats the higher order operator as a perturbative interaction. If one were to try to extrapolate the effective field theory beyond its range of validity, one would find that other physics is needed to describe accurately the higher energy theory. In the QED example, the form of the vacuum polarization changes to a logarithmic function at higher energy.

With this in mind, we can note that there is a variant on our simple model which is much closer to the modern application in quadratic gravity. This
involves a derivative interaction, mimicking how gravity couples to matter proportional to the energy-momentum tensor,
\beq\label{extra}
{\cal L}_{hd}= \frac12 \left[\partial_\mu \phi \partial^\mu \phi - \frac1{M^2} \Box\phi \Box\phi \right] - \kappa(\Box \phi) \chi^\dagger \chi \ \ ,
\eeq
where $\kappa$ is a coupling constant. Now the light $\phi$ field is protected from acquiring a mass by the shift symmetry $\phi \to \phi +c$, and there will be a simple classical limit without having to tune the mass to zero. The analysis of the path integral of this model proceeds similarly to the presentation above, but now the effective low energy interaction from integrating out the heavy field $\eta$ is proportional to $\kappa^2(\Box \chi^\dagger \chi )^2/M^2$. Treated as an effective field theory, this suppressed interaction also does not upset the stability of the theory.

\end{document}